\documentclass{article}

 \usepackage[preprint]{neurips_2019}
\usepackage{wrapfig,lipsum,booktabs}
\usepackage[utf8]{inputenc} % allow utf-8 input
\usepackage[T1]{fontenc}    % use 8-bit T1 fonts
\usepackage{hyperref}   
\hypersetup{
    colorlinks=true,
    linkcolor=blue,
    filecolor=magenta,      
    urlcolor=red,
}% hyperlinks
\usepackage{url}            % simple URL typesetting
\usepackage{booktabs}       % professional-quality tables
\usepackage{amsfonts}       % blackboard math symbols
\usepackage{nicefrac}       % compact symbols for 1/2, etc.
\usepackage{microtype}      % microtypography
\usepackage{algorithm}
\usepackage{algorithmic}
\usepackage{amsmath}
\usepackage{amssymb}
\usepackage{xcolor}
\usepackage{soul}
\usepackage{graphicx}
\usepackage{multirow}
\newcommand{\isEq}[1]{\overset{#1}{\sim}}

\title{Label  Universal Targeted Attack}

\author{%
  Naveed Akhtar*, Mohammad A.~A.~K.~Jalwana*, Mohammed Bennamoun, Ajmal Mian \\
  \small{*The authors contributed equally, and claim joint first authorship. }\\
  Department of Computer Science and Software Engineering\\
  University of Western Australia\\
  35 Stirling Highway, CRAWLEY 6009, WA. \\
  \small{\texttt{\{naveed.akhtar@,mohammad.jalwana@research.,mohammed.bennamoun@,ajmal.mian@\}uwa.edu.au}} \\
}

\begin{document}

\maketitle

\begin{abstract}
We introduce Label Universal Targeted Attack (LUTA) that makes a deep model predict a label of attacker's choice for `any' sample of a given source class with high probability. Our attack stochastically maximizes the log-probability of the target label for the source class with first order gradient optimization, while accounting for the gradient moments. It also suppresses the  leakage of  attack information to the non-source classes for avoiding the attack suspicions. The perturbations resulting from our attack achieve high fooling ratios on the large-scale ImageNet and VGGFace models, and transfer well to the Physical World. Given full control over the perturbation scope in LUTA, we also demonstrate it as a tool for deep model autopsy. The proposed attack reveals interesting perturbation patterns and observations regarding the deep models. 

\end{abstract}

%\vspace{-5mm}
\section{Introduction}
\label{sec:Intro}
%\vspace{-2mm}
Adversarial examples~[1] are carefully manipulated inputs that appear natural to humans but cause deep models to misbehave.     
Recent years have seen multiple methods to generate manipulative signals (i.e.~perturbations) for fooling deep models on individual input samples~[1], [2], [3] or a large number of samples with high probability~[4], [5] - termed `universal'  perturbations.  The former sometimes also  launch `targeted' attacks, where the model ends up predicting a desired target label for the input adversarial example.
The existence of adversarial examples is being widely perceived as a threat to deep learning~[6]. Nevertheless, given appropriate control over the underlying manipulative signal, adversarial examples may also serve as empirical tools for analyzing deep models.

This work introduces a technique to generate manipulative signals that can essentially fool a deep model to confuse `an entire class label' with another label of choice. The resulting Label Universal Targeted Attack (LUTA)\footnote{The source code is provided \href{https://github.com/AsimJalwana/LUTA}{here}. LUTA is intended to be eventually incorporated in public attack libraries, e.g.~\href{https://github.com/bethgelab/foolbox}{foolbox}~[7].} is of high relevance in practical settings. It allows pre-computed perturbations that can change an object's category or a person's identity for a deployed model on-the-fly, where the attacker has also the freedom to choose the target label, and there is no particular constraint over the input. Moreover, the convenient control over the manipulative signal in LUTA encourages the fresh perspective of seeing adversarial examples as model analysis tools.   
Controlling the perturbation scope to individual classes reveals insightful patterns and meaningful information about the classification regions learned by the deep models. %AJMAL: this paragraph does not mention that the attack is source class specific as it minimizes leakage to other source classes. It comes next ... maybe you don't want to emphasize that much?

The proposed LUTA is an iterative algorithm that performs a stochastic gradient-based optimization to maximize the log-probability of the target class prediction for the perturbed source class. It also inhibits fooling of the model on non-source classes to mitigate  suspicions about the attack. 
The algorithm performs careful adaptive learning of the perturbation parameters based on their first and second moments. 
This paper explores three major variants of LUTA. The first two bound the perturbations in  $\ell_{\infty}$ and $\ell_2$ norms, whereas the third allows unbounded perturbations to freely explore the classification regions of the target model. Extensive experiments for fooling VGG-16~[8], ResNet-50~[9], Inception-V3~[10], MobileNet-V2~[11] on ImageNet dataset~[12] and ResNet-50 on large-scale VGG-Face2 dataset~[13] ascertain the effectiveness of our attack.
The attack is also demonstrated in the Physical World.
The unbound LUTA variant is shown to reveal interesting perturbation patterns and insightful observations regarding deep model classification regions.

%\vspace{-3mm}
\section{Prior art}
\label{sec:PA}
%\vspace{-2mm}
Adversarial attacks is currently a highly active research direction. For a comprehensive review, we refer to~[6]. Here, we  discuss the key contributions that relate to this work more closely. 

Szegedy et al.~[1] were the first to report the vulnerability of modern deep learning to adversarial attacks. They showed the possibility of altering images with imperceptible additive perturbations  to  fool deep models. Goodfellow et al.~[2] later proposed the Fast Gradient Sign Method (FGSM) to efficiently estimate such perturbations.  FGSM computes the desired  signal using  the sign of the network's cost function gradient \textit{w.r.t} the input image. The resulting perturbation performs a one-step gradient ascent over the network loss for the input. Instead of a single step, Kurakin et al.~[14] took  multiple small steps for more effective perturbations. They additionally proposed to take steps in the direction that maximizes the prediction  probability of the least-likely class for the image. Madry et al.~[15] noted that the  `projected gradient descent on the negative loss function' strategy adopted by Kurakin et al.~results in highly effective attacks. 
DeepFool~[3] is another popular attack that computes perturbations iteratively by linearizing the model's decision boundaries near the input images. %Similarly, Carlini and Wagner~{\color{red}[CW]} is another instance of currently popular iterative  attacks on deep models.

For the above, the domain of  the computed perturbation is restricted  to a single  image. Moosavi-Dezfooli et al.~[4]  introduced an image-agnostic perturbation to fool a model into misclassifying `any' image. Similar `universal' adversarial perturbations are also computed in [16], [17]. These attacks are non-targeted, i.e.~the adversarial input is allowed to be misclassified into any class. Due to their broader domain, universal perturbations are able to reveal  interesting geometric correlations among the decision boundaries of the deep models~[4], [18]. However, both the perturbation domain and the  model prediction remain unconstrained for the universal perturbations. Manipulative signals are expected to be more revealing with appropriate scoping at those ends. This motivates the need of our label-universal attack that provides control over the source and target labels for model fooling. Such an attack  has also high practical relevance, because it enables the attacker to  conveniently manipulate the semantics learned by a deep model in an unrestricted manner. 

%\vspace{-3mm}
\section{Problem formulation}
\label{sec:PF}
%\vspace{-2mm}
In line with the main stream of research in adversarial attacks, this work considers natural images as the data and model domain. However, the proposed attack is generic under white-box settings.

Let $\boldsymbol\Im \in \mathbb R^{\text d}$  denote the distribution of  natural images, and `$\ell$' be the  label of its random sample ${\bf I}_{\ell} \isEq{\text{rand}} \boldsymbol\Im$. Let $\mathcal C(.)$ be the classifier that  maps $\mathcal C({\bf I}_{\ell}) \rightarrow \ell$ with high probability. We restrict the classifier to be a deep neural network with cross-entropy loss. To fool $\mathcal C(.)$, we seek a perturbation $\boldsymbol{\rho} \in \mathbb R^{\text{d}}$ that satisfies the following constraint:
\begin{align}
\underset{{\bf I}_{\ell} \sim \boldsymbol\Im}{\mathrm{\text P}} \Big( \mathcal C({\bf I}_{\ell} + \boldsymbol \rho) \rightarrow \ell_{\text{target}} : \ell_{\text{target}} \neq \ell \Big) \geq \zeta ~~~\text{s.t.}~~||\boldsymbol\rho||_p \leq \eta,
\label{eq:def}
\end{align}
where `$\ell_{\text{target}}$' is the target label we want $\mathcal C(.)$ to predict for ${\bf I}_{\ell}$ with  probability `$\zeta$' or higher, `$\eta$' controls the $\ell_p$-norm of the perturbation, which is denoted by $||.||_p$. 
In the above constraint, %a complete source class (i.e.~label `$\ell$') forms the domain of $\boldsymbol{\rho}$, implying; 
the same perturbation must fool the classifier on all samples of the source class (labelled `$\ell$') with probability $\geq \zeta$. At the same time, `$\ell_{\text{target}}$' can be any label that is known to $\mathcal C(.)$. 
%$\boldsymbol{\rho}$ aims at changing the label of a whole class (with probability $\geq \delta$) for the classifier.
This formulation inspires the name Label Universal Targeted Attack. % for the fooling caused by the perturbations satisfying   Eq.~(\ref{eq:def}).

%AJMAL: Eqn (1) does not explicitly impose leakage minimization constraint

Allowing `$\ell_{\text{target}}$' to be a random label while  ignoring the label of the input generalizes Eq.~(\ref{eq:def}) to the universal  perturbation constraint~[4]. On the other end, restricting ${\bf I}_{\ell}$ to a single image results in an image-specific targeted attack. In that case, the notion of probability can be ignored. In the spectrum of adversarial attacks forming special cases of Eq.~(\ref{eq:def}), other intermediate choices may include expanding the input domain to a few classes, or using multiple target labels for fooling. Whereas these alternates are not our focus, our algorithm is readily extendable to these cases.

%\vspace{-3mm}
%Degeneration to other perturbations....norms considered.
\section{Computing the perturbation}
\label{sec:Alg}
%\vspace{-2mm}
We compute the perturbations for Label Universal Targeted Attack (LUTA) as shown in  Algorithm~\ref{alg:main}. The abstract concept of the algorithm is intuitive. For a given source class, we compute the desired perturbation by taking small steps over the model's cost surface in the directions that increase the log-probability of the target label for the source class. The directions are computed stochastically, and the steps are only taken in the trusted regions that are governed by the first and (raw) second moment estimates of the directions. While computing a direction, we ensure that it also suppresses the prediction of non-source classes as the target class. To bound the perturbation norm, we keep projecting the accumulated signal to the $\ell_p$-ball of the desired norm at each iteration. The text below  sequentially explains each step of the algorithm in detail. Henceforth, we alternatively refer to the proposed algorithm as LUTA.

\newcommand{\Exp}[1]{\underset{#1}{\mathbb E}}
\begin{algorithm*}[h]
 \caption{Label Universal Targeted Attack}
 \label{alg:main}
 \begin{algorithmic}[1]
 \renewcommand{\algorithmicrequire}{\textbf{Input:}}
 \renewcommand{\algorithmicensure}{\textbf{Output:}}
 \REQUIRE  Classifier $\mathcal C$, source class samples $\mathcal S$, non-source class samples $\overline{\mathcal{S}}$, target label $\ell_{\text{target}}$, perturbation norm $\eta$, mini-batch size $b$,  fooling ratio $\zeta$.
 \ENSURE Targeted label universal perturbation $\boldsymbol{\rho} \in \mathbb R^{\text{d}}$.
 \STATE Initialize $\boldsymbol{\rho}_0$, $\boldsymbol{\upsilon}_0$, $\boldsymbol{\omega}_0$ to zero vectors  in $\mathbb R^{\text{d}}$ and $t = 0$. Set  $\beta_1 = 0.9$, and $\beta_2 = 0.999$.
 %set $\mathcal{P}_s =\{\}$; $\ell_2$-threshold $= \mathbb E\left[ \{ ||\boldsymbol\rho_{i \in \mathcal P}||_2 \}_{i =1}^{|\mathcal P|} \right]$; \\$\mathcal P_n = \mathcal P$ with $\ell_2$-normalized elements.
\WHILE {fooling ratio $< \zeta$} 
\STATE $\mathcal S_s \isEq{\text{rand}} \mathcal S$,~$\mathcal S_o \isEq{\text{rand}} \overline{\mathcal S}$ : $|\mathcal S_s| = |\mathcal S_o| = \frac{b}{2}$  \hspace{11.5mm}$\triangleleft$ {\scriptsize \textit{\textsf{get random samples from the source and other classes}}}
\STATE ${\mathcal S_s} \leftarrow \text{Clip} \left ( \mathcal S_s \ominus \boldsymbol{\rho}_{t} \right)$,~$ {\mathcal S_o} \leftarrow \text{Clip} \left ( \mathcal S_o \ominus \boldsymbol{\rho}_{t} \right)$ \hspace{2.5mm}$\triangleleft$ {\scriptsize \textit{\textsf{perturb and clip samples with the current estimate}}}
\STATE $t \leftarrow  t+1$               \hspace{49.7mm}$\triangleleft$ {\scriptsize \textit{\textsf{increment}}} 
\STATE $\delta \leftarrow \frac{\Exp{{\bf s}_i \in \mathcal S_s}[|| \nabla_{{\bf s}_i} \mathcal J({\bf s}_i, \ell_{\text{target}})||_2 ]}  {\Exp{{\bf s}_i \in \mathcal S_o}[|| \nabla_{{\bf s}_i} \mathcal J({\bf s}_i, \ell)||_2 ]}$ \hspace{25mm}$\triangleleft$ {\scriptsize \textit{\textsf{compute scaling factor for gradient normalization}}}
\STATE $\boldsymbol\xi_t \leftarrow   \frac{1}{2}\left(\Exp{{\bf s}_i \in \mathcal S_s} \big[ \nabla_{{\bf s}_i} \mathcal J({\bf s}_i, \ell_{\text{target}}) \big] + \delta \Exp{{\bf s}_i \in \mathcal S_o} \big[ \nabla_{{\bf s}_i} \mathcal J({\bf s}_i, \ell) \big] \right) $  \hspace{11mm}$\triangleleft$ {\scriptsize \textit{\textsf{compute Expected gradient}}} 
%\STATE $\boldsymbol\xi_n \leftarrow  \boldsymbol\xi / || \boldsymbol\xi||_2$  \hspace{50mm}$\triangleleft$ {\scriptsize \textit{\textsf{normalize}}} 
\STATE $\boldsymbol\upsilon_t \leftarrow \beta_1 \boldsymbol{\upsilon}_{t-1} + (1-\beta_1) \boldsymbol{\xi}_t$  \hspace{24.5mm}$\triangleleft$ {\scriptsize \textit{\textsf{first moment estimate}}} 
\STATE $\boldsymbol{\omega}_t \leftarrow \beta_2 \boldsymbol{\omega}_{t-1} + (1 - \beta_2) (\boldsymbol{\xi}_t\odot \boldsymbol{\xi}_t)$ \hspace{13.7mm}$\triangleleft$ {\scriptsize \textit{\textsf{raw second  moment estimate}}}
\STATE $\boldsymbol{\rho} \leftarrow \frac{\sqrt{1-\beta_2^t}}{1-\beta_1^t}~\text{diag}\left( \text{diag}(\sqrt{\boldsymbol{\omega}_t})^{-1} \boldsymbol{\upsilon}_t \right)$ \hspace{12
mm}$\triangleleft$ {\scriptsize \textit{\textsf{bias corrected moment ratio}}}
%\STATE $ \boldsymbol{\rho} \leftarrow \left( \boldsymbol{\upsilon}_t \sqrt{1- \beta_2^t} \right) \odot \left( {\sqrt{\boldsymbol{\omega}_t} (1-\beta_1^t) } \right)^{-1}$ \hspace{20mm}$\triangleleft$ {\scriptsize \textit{\textsf{intermediate perturbation}}}
\STATE $\boldsymbol{\rho_t} \leftarrow \boldsymbol{\rho}_{t-1} +~\frac{\boldsymbol{\rho}}{|| \boldsymbol{\rho}||_{\infty}}$ \hspace{34mm}$\triangleleft$ {\scriptsize \textit{\textsf{update perturbation}}} %\text{sign} \left( \frac{\boldsymbol{\upsilon}_t \sqrt{1- \beta_2^t}}{\sqrt{\boldsymbol{\omega}_t} (1-\beta_1^t)}  \right)$ \hspace{20mm}$\triangleleft$ {\scriptsize \textit{\textsf{update perturbation}}}
\STATE $\boldsymbol{\rho_t} \leftarrow   \Psi (\boldsymbol{\rho}_t)$ \hspace{45.7mm}$\triangleleft$ {\scriptsize \textit{\textsf{project on $\ell_p$ ball}}}
%\STATE $\boldsymbol\xi = \frac{1}{n} \sum\limits_{i= 1}^n  \nabla \mathcal J_{\mathcal M}({\bf s}_i, \ell_{\text{target}})  $
\ENDWHILE
  \STATE return 
 \end{algorithmic}
 \end{algorithm*}

Due to its white-box nature, LUTA expects the target classifier as one of its inputs. It also requires a set $\mathcal S$ of the source class samples, and a set $\overline{\mathcal S}$ that contains samples of the non-source classes.
Other input parameters include the desired $\ell_p$-norm `$\eta$' of the perturbation, target label `$\ell_{\text{target}}$', mini-batch size `$b$' for  the underlying stochastic optimization, and the desired fooling ratio  `$\zeta$' - defined as the percentage of the source class samples predicted as the target class instances.

We momentarily defer the discussion on  hyper-parameters `$\beta_1$' and `$\beta_2$' on \textit{line 1} of the algorithm. In a given iteration, LUTA first constructs sets  $\mathcal S_s$ and $\mathcal{S}_o$ by randomly sampling the source and non-source classes, respectively. The cardinality of these sets is fixed to `$\frac{b}{2}$' to keep the mini-batch size to  `$b$' (\textit{line 3}). Each element of both sets is then perturbed with the current estimate of the perturbation - operation denoted by symbol  $\ominus$ on \textit{line 4}. The chosen symbol emphasizes that  $\boldsymbol{\rho}_t$ is subtracted in our algorithm from all the samples to perturb them. The `Clip(.)' function clips the perturbed  samples to its valid range, $[0, 255]$ in our case of 8-bit image representation.

\noindent{\bf Lemma 4.1:} \textit{For $\mathcal C(.)$ with cross-entropy cost $\mathcal J(\boldsymbol{\theta}, {\bf s}, \ell)$, the log-probability of ${\bf s}$ classified as `$\ell$' increases in the direction $-\frac{\nabla_{{\bf s}} \mathcal J(\boldsymbol{\theta}, {\bf s},  \ell)}{||\nabla_{{\bf s}} \mathcal J(\boldsymbol{\theta}, {\bf s},  \ell)||_{\infty}}$, where $\boldsymbol{\theta}$ denotes the model parameters\footnote{The model parameters remain fixed throughout, hence we ignore $\boldsymbol{\theta}$ in  Algorithm~\ref{alg:main} and its description.}.}\\
\noindent{\bf Proof:} We can write $\mathcal J(\boldsymbol{\theta}, {\bf s}, \ell) = - \log\left(\text{P}(\ell| {\bf s})\right)$ for $\mathcal C(.)$. Linearizing the cost and inverting the sign, the log-probability maximizes along  $\boldsymbol{\gamma} = {\bf } -\nabla_{{\bf s}} \mathcal J(\boldsymbol{\theta}, {\bf s}, \ell)$. With $||\boldsymbol{\gamma}||_{\infty} = \max_i|\gamma_i|$, $\ell_{\infty}$-normalization re-scales $\boldsymbol{\gamma}$ in the same  direction of increasing $\log\left(\text{P}(\ell | {\bf s})\right)$.

Under Lemma 4.1, LUTA strives to take steps along the cost function's gradient \textit{w.r.t.}~an  input ${\bf s}_i$. Since the domain of ${\bf s}_i$ spans multiple samples in our case, we must take steps along the `Expected' direction of those samples. However, it has to be ensured that the computed direction is not too generic to also cause log-probability rise for the irrelevant (i.e.~non-source class) samples. From the practical view point, perturbations causing samples of `any' class to be misclassified into the target class are less interesting, and can easily raise suspicions. Moreover, they also compromise our control over the perturbation scope, which is not desired.
To refrain from general fooling directions, we  nudge the computed direction such that it also inhibits the fooling of non-source class samples. \textit{Lines 6 and 7} of the algorithm implement these steps as follows. 

On \textit{line 6}, we estimate the ratio between the Expected norms of the source sample gradients and the non-source sample gradients. Notice that we compute the respective gradients using different prediction labels. In the light of Lemma 4.1, $\nabla_{{\bf s}_i} \mathcal J({\bf s}_i, \ell_{\text{target}}): {\bf s}_i \in \mathcal S_s$ gives us the direction (ignoring the negative sign) for fooling a model into predicting label `$\ell_{\text{target}}$' for ${\bf s}_i$, where the sample is from the source class. On the other hand, $\nabla_{{\bf s}_i} \mathcal J({\bf s}_i, \ell): {\bf s}_i \in \mathcal S_o$ provides the direction that improves the model confidence on the correct prediction of ${\bf s}_i$, where the sample is from non-source class. The diverse nature of the computed gradients can result in significant difference between their norms. The scaling factor `$\delta$' on \textit{line 6} is computed  to account for that difference in the subsequent steps.
For the $t^{\text{th}}$ iteration, we compute the Expected gradient $\boldsymbol{\xi}_t$ of our mini-batch on \textit{line 7}.  At this point, it is worth noting that the effective mini-batch for the underlying stochastic optimization in LUTA comprises clipped samples in the set $\mathcal S_s \bigcup \mathcal S_o$. The vector $\boldsymbol{\xi}_t$ is computed as the weighted  average of the Expected gradients of the source and non-source samples. Under the linearity of the Expectation operator and preservation of the vector direction with  scaling, it is straightforward to see that $\boldsymbol{\xi}_t$ encodes the Expected direction to achieve the targeted fooling of the source samples into the label `$\ell_{\text{target}}$', while inhibiting the fooling of non-source samples by increasing their prediction confidence for their correct classes.   

Owing to the diversity of the samples in its mini-batch, LUTA steps in the direction of computed gradient   cautiously. 
On \textit{line 8} and \textit{line 9}, it respectively estimates the first and the raw second moment (i.e.~un-centered variance) of the computed gradient using exponential moving averages. The hyper-parameters `$\beta_1$' and `$\beta_2$' decide the decay rates of these averages, whereas $\odot$ denotes the Hadamard product. The use of moving averages as the moment estimates in LUTA is inspired by the Adam algorithm~[19] that efficiently performs stochastic optimization. However, instead of using the moving averages of gradients to update the parameters (i.e.~model weights) as in [19], we compute those for the Expected gradient and capitalize on the directions for perturbation estimation. Nevertheless, due to the similar physical significance of the hyper-parameters  $\beta_1,~\beta_2 \in [0,1)$ in LUTA and Adam, the performance of both algorithms largely remains insensitive to small changes to the values of these parameters. Following~[19], we fix  $\beta_1 = 0.9, \beta_2 = 0.999$ (\textit{line 1}). We refer to [19] for further details on the choice of these values for the gradient based stochastic optimization. 

The gradient moment estimates in LUTA are exploited in stepping along the cost surface. Effectiveness of the moments as stepping guides for stochastic optimization is already well-established~[19], [20]. Briefly ignoring the expression for $\boldsymbol{\rho}$ on \textit{line~10} of the algorithm, we compute this guide as the ratio between the moment estimates    $\frac{\boldsymbol{\upsilon}_t}{\sqrt{\boldsymbol{\omega}_t}}$, where the square-root accounts for $\boldsymbol{\omega}_t$ representing the `second' moment. Note that, we slightly abuse the notation here as both values are vectors. On \textit{line~10}, we use the mathematically correct expression, where diag(.) converts a vector into a diagonal matrix, or a diagonal matrix into a vector, and the inverse is performed element-wise. Another improvement in \textit{line~10} is that we use the  `bias-corrected' ratio of the moment estimates instead. Moving averages are known to get heavily biased at early iterations. This becomes a concern when the algorithm can benefit from well-estimated initial points. In our experiments (\S \ref{sec:Exp}), we also use LUTA in that manner. Hence, bias-correction is accounted for in our technique. We provide a detailed derivation to arrive at the expression on \textit{line~10} of Algorithm~\ref{alg:main} in \S A-1 of the supplementary material.  

Let us compactly write $\boldsymbol{\rho} = \frac{\widetilde{\boldsymbol{\upsilon}}_t}{\sqrt{\widetilde{\boldsymbol{\omega}}_t}}$, where tilde indicates the bias-corrected vectors. It is easy to see that for a large  second moment estimate $\widetilde{\boldsymbol{\omega}}$, $\boldsymbol{\rho}$ shrinks. This is desirable because we eventually take a step along $\boldsymbol{\rho}$, and a  smaller step is preferable along the components that have larger variance.
The perturbation update step on \textit{line 11} of the algorithm further restricts $\boldsymbol{\rho}$ to unit $\ell_{\infty}$-norm.
To an extent, this relates to computing the gradient's sign in FGSM~[2]. However, most coefficients of $\boldsymbol{\rho}$ get restricted to smaller values in our case instead of $\pm 1$. As a side remark, we note that simply computing the sign of $\boldsymbol\rho$ for perturbation update eventually nullifies the advantages of the second moment estimate due to the squared terms.   
The $\ell_{\infty}$ normalization is able to preserve the required direction in our case, while taking full advantage of the second moment estimate.  

\noindent{\bf LUTA variants:}~As seen in Algorithm~\ref{alg:main}, LUTA accumulates the signals computed at each iteration.
To restrict the norm of the accumulated perturbation, $\ell_p$-ball projection is used. The use of different types of balls results in different variants of the algorithm. For the  $\ell_{\infty}$-ball projection,  we implement $\Psi (\boldsymbol{\rho}_t) = \text{sign}(\boldsymbol\rho_t) \odot \text{min}\left(\text{abs}(\boldsymbol\rho_t), \eta\right)$ on \textit{line 12}. In the case of $\ell_2$-ball projection, we use $\Psi (\boldsymbol{\rho}_t) = \text{min}\left(1, \frac{\eta}{||\boldsymbol{\rho}_t||_2}\right) \boldsymbol\rho_t$. These  projections respectively bound the $\ell_{\infty}$ and $\ell_2$ norms of the perturbations. We bound these norms to reduce the perturbation's perceptibility, which is in line with the existing literature. However, we also employ a variant in which $\Psi (\boldsymbol{\rho}_t) = \mathcal I(\boldsymbol{\rho_t})$, where $\mathcal I(.)$ is the identity mapping. We refer to this particular variant as LUTA-U, for  the `Unbounded' perturbation norm. In contrast to the typical use of  perturbations in adversarial attacks, we employ LUTA-U perturbations to explore the classification regions of the target model without restricting their norm.
Owing to the `label-universality' of the perturbations, LUTA-U exploration promises to reveal interesting information regarding the  classification regions of the deep models.

\vspace{-3mm}
\section{Evaluation}
\label{sec:Exp}
\vspace{-2mm}
We evaluate the proposed LUTA as an attack in \S~\ref{sec:att} and as an exploration tool in \S~\ref{sec:LUTAU}. For the latter, the unbounded version (LUTA-U) is used. 

\vspace{-3mm}
\subsection{LUTA as attack}
\label{sec:att}
\vspace{-2mm}
\paragraph{Setup:} We first demonstrate the success of label-universal targeted fooling under LUTA by attacking  VGG-16~[8], ResNet-50~[9], Inception-V3~[10] and MobileNet-V2~[11]  trained on ImageNet dataset~[12]. We use  \href{https://keras.io/applications/}{Keras} provided public models, where selection of the networks is based on their established performance and diversity.  We use the training set of ILSVRC2012 for perturbation estimation, whereas the validation set of this data  (50 samples per class) is used as our test set.  For the non-source classes, we only use the correctly classified samples during training, with a lower bound of $60\%$ on the prediction confidence. 
This filtration is  performed for computational purpose. It still ensures useful gradient directions  with fewer non-source samples. We do not filter the source class data. We compute a perturbation using a two step strategy. First, we alter Algorithm~\ref{alg:main} to disregard the non-source class data. This is achieved by  replacing the non-source class set $\overline{\mathcal S}$ with the source class set $\mathcal S$ and using `$\ell_{\text{target}}$' instead of `$\ell$' for the  gradient computation. In the second step, we initialize LUTA with the perturbation computed in the first step.
This procedure is also adopted for computational gain under better initialization. In the first step, we let the algorithm run for 100 iterations, while $\zeta$ is set to $80\%$ in the second step. We ensure at least 100 additional iterations in the second step. We empirically set `$b$' to 64 for the first step and 128 for the second. In the text to follow, we discuss the setup details only when those are different from what is  described here.     

Besides fooling the ImageNet models, we also attack VGGFace model~[13] (ResNet-50 architecture) trained on the large-scale VGG-Face2 dataset~[13]. In our experiments, Keras provided model weights are used that are converted from the original Caffe implementation. We use the training set of VGG-Face2 and crop the faces using the bounding box   \href{http://www.robots.ox.ac.uk/~vgg/data/vgg_face2/meta_infor.html}{meta data}. Random 50 images for an identity are used as the test set, while the remaining images are used for perturbation estimation. 

\noindent{\bf Fooling ImageNet models:}~We randomly choose ten source classes from  ImageNet and make another random selection of ten target labels, resulting in ten label transforming (i.e.~fooling) experiments for a single model. Both $\ell_{\infty}$ and $\ell_2$-norm bounded perturbations are then  considered, letting  $\eta=15$ and $4,500$ respectively. The `$\eta$' values are chosen based on perturbation perceptibility. %As will be seen shortly, the perturbations remain imperceptible to quasi-imperceptible with the chosen `$\eta$' values, while achieving high fooling ratios. %

We summarize the results in Table~\ref{tab:success}.  Note that the reported fooling ratios are on `test' data that is previously unseen by both the targeted model and our algorithm. Successful fooling of the large-scale models is apparent from the Table. The last column reports `Leakage', which is defined as the average fooling ratio of the non-source classes into the target label. Hence, lower Leakage values are more desirable.  
It is worth mentioning that in a separate experiment where we alter our algorithm so that it does not suppress  fooling of the non-source classes, a significant rise in the Leakage  was observed. We provide results of that experiment in \S A-2 of the supplementary material. 
Table~\ref{tab:success} caption provides the label information for the source~$\rightarrow$~target transformation employing the commonly used nouns. We refer to  \S A-3 of the supplementary material for the exact labels and original WordNet IDs of the ImageNet dataset.

\begin{table}[t]
\centering
\setlength{\tabcolsep}{5.7pt}
{\scriptsize
\begin{tabular}{c|l|c|c|c|c|c|c|c|c|c|c||c|c}
  \hline  
   {\bf Bound} & {\bf Model} & {\bf T$_1$} & {\bf T$_2$} & {\bf T$_3$} & {\bf T$_4$} & {\bf T$_5$} & {\bf T$_6$} & {\bf T$_7$} & {\bf T$_8$} & {\bf T$_9$} & {\bf T$_{10}$} &  {\bf Avg.} & {\bf Leak.} \\ \hline \hline
   \multirow{ 4}{*}{$\ell_{\infty}$-norm} & VGG-16~[8]           & 92 & 76 & 80 & 74 & 82 & 78 & 82 & 80 & 74 & 88 &  80.6$\pm$5.8 & 29.9   \\
                                             & ResNet-50~[9]     & 92 & 78 & 80 & 72 & 76 & 84 & 78 & 76 & 82 & 78 &  79.6$\pm$5.4 &  31.1 \\
                                             & Inception-V3~[10] & 84 & 60 & 70 & 60 & 68 & 90 & 68 & 62 & 72 & 76 &  71.0$\pm$9.9 &   24.1 \\
                                             & MobileNet-V2~[11]        & 92 & 94 & 88 & 78 & 88 & 86 & 74 & 86 & 84 & 94 &  86.4$\pm$6.5 &  37.1 \\ \hline \hline
   \multirow{ 4}{*}{$\ell_{2}$-norm}      & VGG-16~[8]           & 90 & 84 & 80 & 84 & 94 & 86 & 82 & 92 & 86 & 96 &  87.4$\pm$5.3 & 30.4\\
                                             & ResNet-50~[9]     & 96 & 94 & 88 & 84 & 90 & 86 & 86 & 94 & 90 & 90 &  89.8$\pm$3.9 & 38.0\\
                                             & Inception-V3~[10] & 86 & 68 & 62 & 62 & 74 & 72 & 74 & 68 & 66 & 76 &  70.8$\pm$7.2 & 45.6\\
                                             & MobileNet-V2~[11] & 94 & 98 & 92 & 76 & 94 & 92 & 76 & 92 & 92 & 96 &  90.2$\pm$7.7 & 56.0\\ \hline

  \hline
\end{tabular}}
\caption{{\small Fooling ratios (\%) with $\eta = 15$ for $\ell_{\infty}$ and $4,500$ for $\ell_2$-norm bounded label-universal perturbations for ImageNet models. The label transformations are as follows. T$_1$: Airship $\rightarrow$ School Bus, T$_2$: Ostrich $\rightarrow$ Zebra, T$_3$: Lion $\rightarrow$ Orangutang, T$_4$: Bustard $\rightarrow$ Camel, T$_5$: Jelly Fish $\rightarrow$ Killer Whale, T$_6$: Life Boat $\rightarrow$ White Shark, T$_7$: Scoreboard $\rightarrow$ Freight Car, T$_8$: Pickelhaube $\rightarrow$ Stupa, T$_9$: Space Shuttle $\rightarrow$ Steam Locomotive, T$_{10}$: Rapeseed $\rightarrow$ Butterfly. Leakage (last column) is the average fooling of non-source classes into the target label.}}
\vspace{-5mm}
\label{tab:success}
\end{table}

\begin{figure}
    \centering
    \includegraphics[height= 3in]{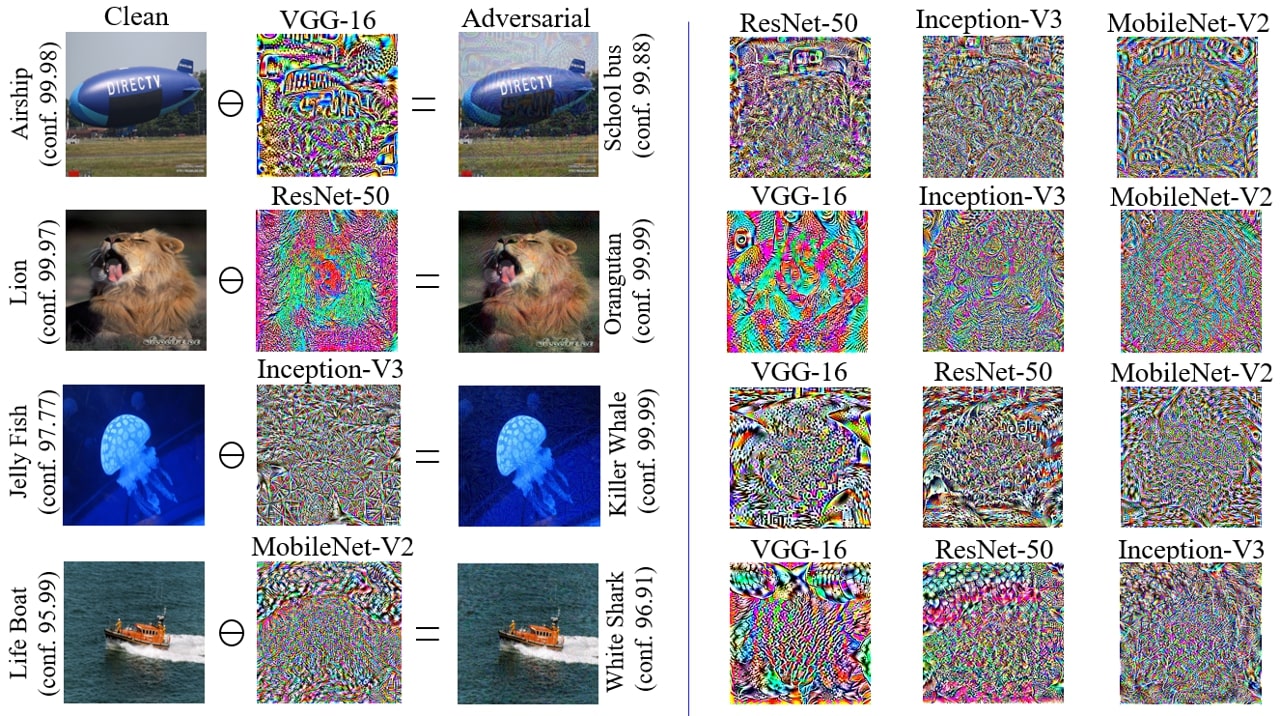}
    \caption{Representative  perturbations and adversarial images for $\ell_{\infty}$-bounded case, $\eta=15$. Each row shows perturbations for the same source $\rightarrow$ target fooling for the mentioned networks. An adversarial example for a model is also shown for reference (on left),  reporting the model confidence on the target label. Following [1],  the perturbations are visualized by 10x magnification, shifted by 128 and clamped to 0-255. Refer to \S A-4 of the supplementary document for more examples.}
    \label{fig:illus1}
    \vspace{-5mm}
\end{figure}

In Fig.~\ref{fig:illus1}, we show perturbations for representative label foolings. The figure also presents a sample adversarial example for each network. In our experiments, it was frequently observed that the models show high confidence on the adversarial samples, as it is also clear from the figure. We provide further images for both $\ell_{\infty}$ and $\ell_2$-norm perturbations in \S A-4 of the supplementary material. From the  images, we can see that the perturbations are often not easy to perceive by the Human visual system. It is emphasized that this perceptibility and the fooling ratio in Table~\ref{tab:success} is based on the selected `$\eta$' values. Allowing larger `$\eta$'  results in even higher fooling ratios at the cost of larger perceptibility.

\begin{table}[b]
\vspace{-2mm}
    \centering
    \begin{tabular}{c|c|c|c|c|c|c||c|c|c|c|c|c|c}\hline
\multicolumn{7}{c||}{$\ell_{\infty}$-norm bounded}&
\multicolumn{7}{c}{$\ell_{2}$-norm bounded} \\
\hline \hline
  F$_1$ & F$_2$ & F$_3$ & F$_4$ & F$_5$ & Avg. & Leak. & F$_1$ & F$_2$ & F$_3$ & F$_4$ & F$_5$ & Avg. & Leak. \\ \hline
  88 & 76 & 74 & 86 & 84 & 81.6$\pm$6.2 & 1.9 & 76 & 80 & 78 & 76 & 84 & 78.8$\pm$3.3 & 1.8 \\ \hline
    \end{tabular}
    \caption{Switching face identities for VGGFace model on test set with LUTA (\% fooling): The switched identities in the original dataset are, F$_1$: n000234$\rightarrow$ n008779, F$_2$: n000282 $\rightarrow$ n006494, F$_3$: n000314 $\rightarrow$ n007087, F$_4$: n000558 $\rightarrow$ n001800,
    F$_5$: n005814 $\rightarrow$ n006402. The $\ell_{\infty}$ and $\ell_2$-norms of the perturbation are upper bounded to 15 and 4,500 respectively.}
    \label{tab:face}
    \vspace{-4mm}
\end{table}

\begin{figure}[t]
    \centering
    \includegraphics[width = \textwidth]{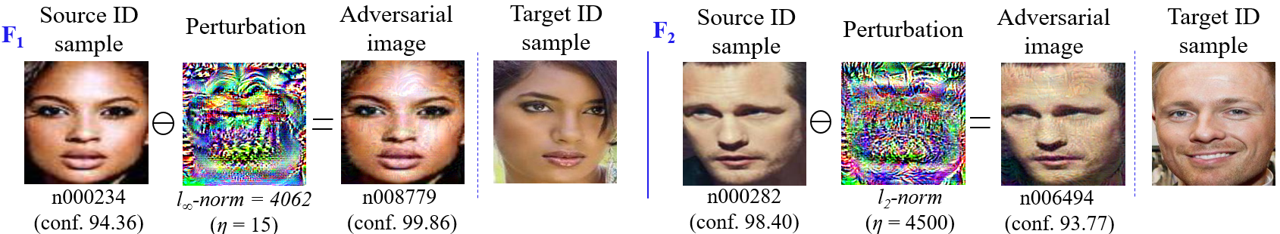}
    \vspace{-2mm}
    \caption{Representative face ID switching examples for VGGFace model. Sample clean target ID image is  provided for reference. Same setup as Table~\ref{tab:face} is used. Perturbation visualization follows [1].}
    \label{fig:face}
    \vspace{-3mm}
\end{figure}

%\vspace{-5mm}
{\bf Fooling VGGFace model:}
We also test our algorithm for switching face identities in the large-scale VGGFace model~[13]. Table~\ref{tab:face} reports the results on five identity switches that are randomly chosen from the  VGG-Face2 dataset. Considering the variety of  expression, appearance, ambient  conditions etc. for a given subject in VGG-Face2, the results in Table~\ref{tab:face} imply that LUTA enables an attacker to change their identity on-the-fly with high probability,  without worrying about the image capturing conditions. Moreover, leakage of the  target label to the non-source classes also remains remarkably low. We conjecture that this happens because the target objects (i.e.~faces) occupy major regions of the images in the dataset, which mitigates the influence of identity-irrelevant information in   perturbation estimation, resulting in a more specific manipulation of the source to target conversion. Figure~\ref{fig:face} illustrates representative adversarial examples resulting from LUTA for the face ID switches.   
Further images can also be found in \S A-5 of the supplementary material. 
The results demonstrate successful identity switching on unseen images by LUTA. 

%\vspace{-5mm}
\subsection{LUTA-U as network autopsy tool} %AJMAL: as network autopsy tool
\label{sec:LUTAU}
%\vspace{-2mm}

\begin{wrapfigure}{r}{0.5\textwidth}
\vspace{-5mm}
  \begin{center}
    \includegraphics[width=0.48\textwidth]{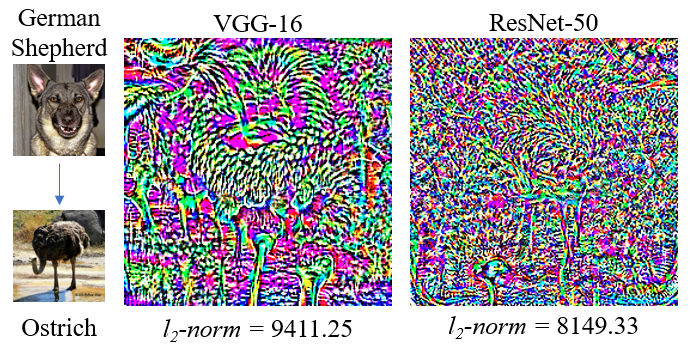}
  \end{center}
  \vspace{-3mm}
  \caption{Patterns emergence with LUTA-U achieving 100\% test accuracy for German Shephered $\rightarrow$ Ostrich. $\ell_2$-norms of perturbations are given. Clean samples are shown for reference.  }
  \label{fig:LUTAU-1}
  \vspace{-3mm}
\end{wrapfigure}

Keeping aside the success of LUTA as an attack, it is intriguing to investigate the patterns that eventually change the semantics of a whole class for a network.
For that, we let LUTA-U run to achieve 100\% test accuracy and observe the perturbation patterns. We notice a repetition of the characteristic visual features of the target class in the perturbations thus created, see  Fig.~\ref{fig:LUTAU-1}. 
Another observation we make is that multiple runs of LUTA lead to different perturbations, nevertheless, those perturbations preserve the characteristic features of the target label.
We refer  to \S A-6 of the supplementary material for the corroborating visualizations.
Besides advancing the proposition that perturbations with broader input domain are able to exploit geometric correlations between the decision boundaries of the classifier~[4], these observations also foretell  (possibly) non-optimization based targeted fooling techniques in the future, where salient visual features of the target class may be cheaply embedded in the adversarial images.

Another interesting use of LUTA-U is in exploring the  classification regions induced by the deep models. 
We employ MobileNet-V2~[11], and let LUTA-U achieve $95\%$ fooling rate on the training samples in each experiment. We choose five ImageNet classes from Table~\ref{tab:success} and convert their labels into each other. We keep the number of training samples the same for each class, i.e.~965 as allowed by the dataset. In our experiment, the perturbation vector's $\ell_2$-norm is used as the representative distance covered by the source class samples to cross over and stay in the  target class region. Experiments are repeated three times and the mean distances are reported in Table~\ref{tab:distance}. Interestingly, the differences between the distances for $A \rightarrow B$ and $B \rightarrow A$ are significant.
On the other hand, we can see  particularly lower values for `Airship' and larger values for `School Bus' for all transformations. These observations are explainable under the hypothesis that $w.r.t.$~the remaining classes, the classification region for `Airship' is more like a blob in the high dimensional space that lets majority of the samples in it move (due to perturbation) more coherently towards other class regions. On the other end, `School Bus' occupies a relatively flat but well-spread  region that is farther from `Space Shuttle' as compared to e.g.~`Life Boat'. 

%AJMAL: in the table below, the values are proportional to the number of pixels (are they?). would averaging them over pixels make more sense? what 
\begin{table}[t]
    \centering
    {\scriptsize
    \begin{tabular}{l||c|c|c|c|c} \hline
         {\bf Target} $\rightarrow$  & Space Shuttle & Steam Locomotive & Airship & School Bus & Life Boat\\
         {\bf Source} $\downarrow$ & &  & & &\\ \hline\hline
         Space Shuttle & - & 4364.4$\pm$81.1  & 4118.3$\pm$74.5 & 4679.4$\pm$179.5 & 5039.1$\pm$230.7 \\
         Steam Locomotive & 5406.8$\pm$57.7 & - & 4954.7$\pm$56.5 & 5845.2$\pm$300.4 & 5680.2$\pm$40.0 \\
         Airship & 3586.4$\pm$59.4 & 3992.7$\pm$291.1 & - & 3929.5$\pm$50.4 & 3937.8$\pm$33.7 \\
         School Bus & 7448.8$\pm$200.9 & 6322.8$\pm$89.5 & 6586.8$\pm$165.1 & - & 5976.5$\pm$112.1 \\
         Life Boat & 5290.4$\pm$43.1 & 5173.0$\pm$71.8 & 5121.5$\pm$154.1 & 5690.9$\pm$47.4 & - \\ \hline
          
    \end{tabular}}
    \caption{Average $\ell_2$-norms of the perturbations to achieve $95\%$ fooling on MobileNet-V2 [11].}
    \label{tab:distance}
    \vspace{-5mm}
\end{table}

\begin{figure}[t]
    \centering
    \includegraphics[width = 5in]{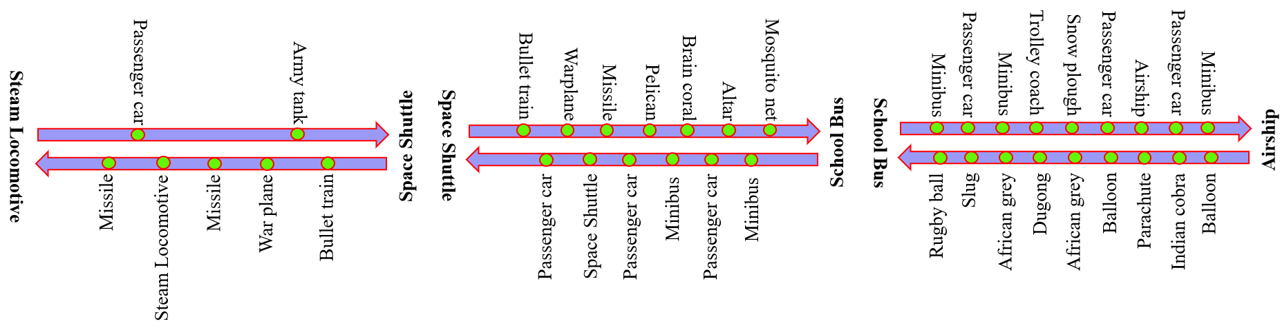}
    \caption{Max-label hopping during transformations using LUTA-U. Setup of Table~\ref{tab:distance} is employed.}
    \label{fig:hops}
    \vspace{-2mm}
\end{figure}

LUTA makes the source class samples collectively move towards the target class region of a model with perturbations.
Hence, LUTA-U iterations also provide a unique opportunity to examine this migration through the classification regions. For the Table~\ref{tab:distance} experiment, we monitor the top-1 predictions during the iterations and record the maximally predicted labels (excluding the source label) during training. In Fig.~\ref{fig:hops}, we show this information as `max-label hopping' for six representative transformations. 
The acute observer will notice that both Table~\ref{tab:distance} and Fig.~\ref{fig:hops} consider `transportation means' as the source and target classes.
This is done intentionally to illustrate the clustering of model  classification regions for semantically similar classes. 
Notice in Fig.~\ref{fig:hops}, the  hopping mostly involves intermediate classes related to transportation/carriage means. Exceptions occur when `School Bus' is the target class. This confirms our hypothesis that this class has a well-spread region. Consequently, it attracts a variety of intermediate labels as the target when perturbed, including those that live (relatively) far from its main cluster.  

Our analysis only scratches the surface of the model exploration and exploitation possibilities enabled by LUTA, promising many interesting future research directions to which the community is invited.     

%\vspace{-3mm}
\subsection{Physical World attack}
\label{sec:Phy}
\vspace{-2mm}
Label universal targeted attacks have serious implications if they transfer well to the Physical World. To evaluate LUTA as the Physical World attack, we observe model label prediction on a live webcam stream of the printed adversarial images. No enhancement/transformation is applied other than color printing the adversarial images. This setup is considerably more challenging than e.g.~fooling on the digital scans of  printed adversarial images~[14]. Despite that, LUTA perturbations are found  surprisingly effective for label-universal fooling in the Physical World.  The exact details of our experiments are provided in \S A-7 of the 
supplementary material. We also provide a video \href{https://youtu.be/-H9ufWlvS44}{here}, capturing the live streaming examples.

%The objective of the label-universal targeted attacks appears too challenging to transfer well to the Physical World. However, the proposed LUTA is able to  demonstrate surprisingly good results for fooling in the Physical World. We demonstrate this by live label prediction of a webcam stream for  printed clean images and their adversarial counterparts obtained with LUTA. Due to the space restriction, details of the experiments are {\color{red}provided in \S A-7 of the supplementary material}, and a video is also included. {\color{red} A sentence on the punch-line/finding.} 

%\vspace{-3mm}
\subsection{Hyper-parameters and training time}
\label{sec:param}
%\vspace{-2mm}
In Algorithm~\ref{alg:main}, the desired fooling ratio `$\xi$' controls the total number of iterations, given fixed mini-batch size `$b$' and `$\eta$'. Also, the mini-batch size plays its typical role in the underlying
\begin{wrapfigure}{r}{0.68\textwidth}
\vspace{-5mm}
  \begin{center}
    \includegraphics[width=0.68\textwidth]{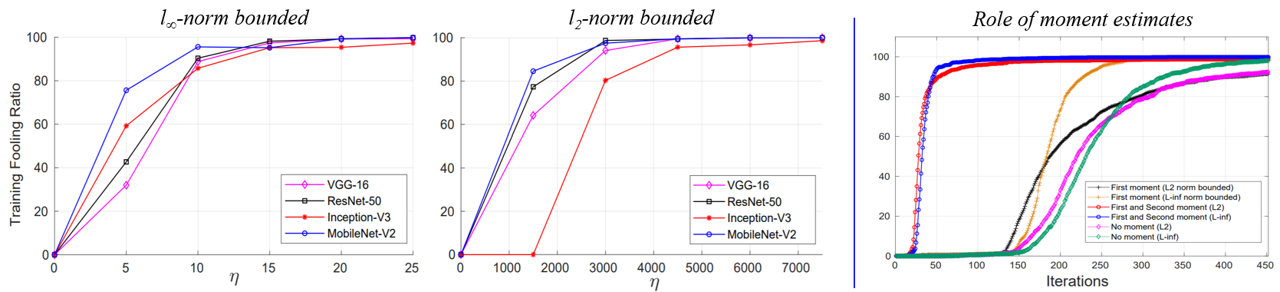}
  \end{center}
  \vspace{-3mm}
  \caption{Effects of varying `$\eta$' on fooling ratio (left). Efficacy of moments in optimization (right), $\eta =$ 15 and 4500 for $\ell_{\infty}$ and  $\ell_2$.}
  \label{fig:param}
  %\vspace{-2mm}
\end{wrapfigure}
stochastic optimization problem. 
Hence, we mainly focus on the parameter `$\eta$' in this section.     
Fig.~\ref{fig:param}~(left) shows the effects of varying `$\eta$' on the fooling ratios for the considered four ImageNet models for both $\ell_{\infty}$ and $\ell_2$-norm bounded perturbations. Only the values of T$_1$ are included for clarity, as other transformations show qualitatively similar behavior. Here, we cut-off the training after 200 iterations. The rise in fooling ratio with larger `$\eta$' is apparent. On average, 100 iterations of our Python 3 LUTA implementation requires 18.8, 20.9, 33.6 and 19.5 minutes for VGG-16, ResNet-50, Inception-V3 and MobileNet-V2 on NVIDIA Titan Xp GPU with 12 GB RAM using `$b = 128$'. Fig.~\ref{fig:param}~(right) also illustrates the role of the first and second moments in achieving the desired fooling rates more efficiently. For clarity, we show it for T$_1$ for MobileNet-V2. Similar qualitative behavior was observed in all our experiments. It is apparent that both moments significantly improve the efficiency of LUTA by estimating the desired perturbation in fewer iterations. We use $\xi = 99\%$ allowing at most 450 iterations. 
%AJMAL: check the "We show it for T$_1$ for .... " sentence

\vspace{-4mm}
\section{Conclusion}
\label{sec:conc}
\vspace{-3mm}
We present the first of its kind attack that changes the label of a whole class into another label of choice. Our white-box attack computes perturbations based on the samples of the source and non-source classes, while stepping in the directions that are guided by the first two moments of the computed gradients. The estimated perturbations are found to be  effective for fooling large-scale ImageNet and VGGFace models, while remaining largely imperceptible. We also show that the label-universal perturbations are able to transfer well to the Physical World. The proposed attack is additionally demonstrated to be an effective tool to empirically explore the classification regions of deep models, revealing insightful modelling details.  Moreover, LUTA perturbations exhibit interesting target label patterns, which opens  possibilities for their blackbox extensions.

%it is demonstrated to reveal interesting patterns in the computed perturbations. % Both as an attack and exploration tool, LUTA is found to be highly relevant.  
%AJMA: change the last sentence to something like this ... the perturbations computed by LUTA reveal interesting patterns specific to the target class and hence opening opportunities for launching blackbox attacks.

\section*{Acknowledgement}
This  work  is  supported  by  Australian  Research  Council Grant  ARC  DP19010244.   The  GPU  used  for  this  work  was donated by NVIDIA Corporation.

%\newpage
\section*{References}
%  \verb+small+ (9 point)
% when listing the references. {\bf Remember that you can use more than eight
%   pages as long as the additional pages contain \emph{only} cited references.}
% \medskip
\small
[1] Szegedy, C., Zaremba, W., Sutskever, I., Bruna, J., Erhan, D., Goodfellow, I. and Fergus, R., 2013. Intriguing properties of neural networks. arXiv preprint arXiv:1312.6199.

[2] Goodfellow, I.J., Shlens, J. and Szegedy, C., 2014. Explaining and harnessing adversarial examples. arXiv preprint arXiv:1412.6572.

[3] Moosavi-Dezfooli, S.M., Fawzi, A. and Frossard, P., 2016. Deepfool: a simple and accurate method to fool deep neural networks. In Proc. IEEE CVPR (pp. 2574-2582).change

[4] Moosavi-Dezfooli, S.M., Fawzi, A., Fawzi, O. and Frossard, P., 2017. Universal adversarial perturbations. In Proc. IEEE CVPR (pp. 1765-1773).

[5] Reddy Mopuri, K., Ojha, U., Garg, U. and Venkatesh Babu, R., 2018. NAG: Network for adversary generation. In Proc. IEEE CVPR (pp. 742-751).

[6] Akhtar, N. and Mian, A., 2018. Threat of adversarial attacks on deep learning in computer vision: A survey. IEEE Access, 6, pp.14410-14430.

[7] Rauber, J., Wieland, B., and Matthias, B., 2017.  Foolbox: A Python toolbox to benchmark the robustness of machine learning models. arXiv preprint arXiv:1707.04131.

[8] Simonyan, K. and Zisserman, A., 2014. Very deep convolutional networks for large-scale image recognition. arXiv preprint arXiv:1409.1556.

[9] He, K., Zhang, X., Ren, S. and Sun, J., 2016. Deep residual learning for image recognition. In Proceedings of the IEEE conference on computer vision and pattern recognition (pp. 770-778).

[10] Szegedy, C., Vanhoucke, V., Ioffe, S., Shlens, J. and Wojna, Z., 2016. Rethinking the inception architecture for computer vision. In Proceedings of the IEEE conference on computer vision and pattern recognition (pp. 2818-2826).

[11] Sandler, M., Howard, A., Zhu, M., Zhmoginov, A. and Chen, L.C., 2018. Mobilenetv2: Inverted residuals and linear bottlenecks. In Proceedings of the IEEE Conference on Computer Vision and Pattern Recognition (pp. 4510-4520).

%[7] Moosavi-Dezfooli, S.M., Fawzi, A., Fawzi, O., Frossard, P. and Soatto, S., 2017. Analysis of universal adversarial perturbations. arXiv preprint arXiv:1705.09554.

[12] Deng, J., Dong, W., Socher, R., Li, L.J., Li, K. and Fei-Fei, L., 2009, June. Imagenet: A large-scale hierarchical image database. In 2009 IEEE conference on computer vision and pattern recognition (pp. 248-255).

[13] Cao, Q., Shen, L., Xie, W., Parkhi, O.M. and Zisserman, A., 2018, May. Vggface2: A dataset for recognising faces across pose and age. In 2018 13th IEEE International Conference on Automatic Face \& Gesture Recognition (FG 2018) (pp. 67-74). IEEE.

[14] Kurakin, A., Goodfellow, I. and Bengio, S., 2016. Adversarial examples in the physical world. arXiv preprint arXiv:1607.02533.

[15] Madry, A., Makelov, A., Schmidt, L., Tsipras, D. and Vladu, A., 2017. Towards deep learning models resistant to adversarial attacks. arXiv preprint arXiv:1706.06083.

[16] Khrulkov, V. and Oseledets, I., 2018. Art of singular vectors and universal adversarial perturbations. In Proceedings of the IEEE Conference on Computer Vision and Pattern Recognition (pp. 8562-8570).

[17] Mopuri, K.R., Garg, U., and Radhakrishnan, V.B. 2017. Fast Feature Fool: A data independent approach to universal adversarial perturbations. arXiv preprint arXiv:1707.05572.

[18] Moosavi-Dezfooli, S.M., Fawzi, A., Fawzi, O., Frossard, P. and Soatto, S., 2017. Analysis of universal adversarial perturbations. arXiv preprint arXiv:1705.09554.

[19] Kingma, D.P. and Ba, J., 2014. Adam: A method for stochastic optimization. arXiv preprint arXiv:1412.6980.

[20] Tielman, T. and Hinton, G. Lecture 6.5 - RMSProp, COURSERA: Neural Networks for Machine Learning, Teachnical Report, 2012.

\newpage
\begin{center}
    \large{\bf Supplementary Material}\\
    \vspace{2mm}
    \small{(Label Universal Targeted Attack)}
\end{center}
\vspace{5mm}

\noindent{\bf A-1: Computing the bias corrected moments ratio}
\vspace{3mm}

To derive the expression for the bias corrected moment ratio in Algorithm~1, we first focus on the moving average expression of $\boldsymbol{\upsilon}_t$:

In $\boldsymbol{\upsilon}_t = \beta \boldsymbol{\upsilon}_{t-1} + (1 - \beta) \boldsymbol{\xi}_t$, we ignore the subscript of `$\beta_1$' for clarity. We can write
\
\\
\
at t = 1:~~$\boldsymbol{\upsilon}_1 = (1- \beta) \boldsymbol{\xi}_1$,
\
\\
\
at t = 2:~~$\boldsymbol{\upsilon}_2 = \beta \boldsymbol{\upsilon}_1 + (1 - \beta) \boldsymbol{\xi}_2 = (1 - \beta) (\beta^1 \boldsymbol{\xi}_1 + \beta^0 \boldsymbol{\xi}_2)$,
\
\\
\
at t = 3:~~$\boldsymbol{\upsilon}_3 = \beta \boldsymbol{\upsilon}_2 + (1 - \beta) \boldsymbol{\xi}_3 = (1 - \beta) (\beta^2 \boldsymbol{\xi}_1 + \beta^1 \boldsymbol{\xi}_2 + \beta^0 \boldsymbol{\xi}_3)$, resulting in the expression:
\begin{align}
    \boldsymbol{\upsilon}_t = (1-\beta) \sum\limits_{i = 1}^t \beta^{t-i} \boldsymbol{\xi}_i.
\label{eq:red}
\end{align}

Using Eq.~(\ref{eq:red}), we can relate the Expected value of $\boldsymbol{\upsilon}_t$ to the Expected value of the true first moment $\boldsymbol{\xi}_t$ as follows:
\begin{align}
    \mathbb E [\boldsymbol{\upsilon}_t] = \mathbb E \left[(1-\beta) \sum\limits_{i = 1}^t \beta^{t-i} \boldsymbol{\xi}_i \right] =  \mathbb E [\boldsymbol \xi_t]~(1-\beta) \sum\limits_{i = 1}^t \beta^{t-1} + \epsilon,
    \label{eq:full}
\end{align}
where $\epsilon \rightarrow 0$ for the value of `$\beta$' assigning very low weights to the more distant time stamps in the past (e.g.~$\beta \geq 0.9$).  Ignoring `$\epsilon$', the remaining expression gets simplified to:
\begin{align}
    \mathbb E [\boldsymbol{\upsilon}_t] = \mathbb E [\boldsymbol{\xi}_t] (1-\beta^t).
    \label{eq:exp}
\end{align}
Simplification of Eq.~(\ref{eq:full}) to Eq.~(\ref{eq:exp}) is verifiable by choosing a small value of `$t$' and expanding the former.
In Eq.~(\ref{eq:exp}), the term $(1- \beta^t)$ causes a bias for a larger $\beta \in [0,1)$ and smaller $t$, which is especially true for the early iterations of the algorithm. Hence, to account for the bias, $ \widetilde{\boldsymbol{\upsilon}}_t =  \frac{\boldsymbol{\upsilon}_t}{(1-\beta_1)^t}$ must be used instead of directly employing  $\boldsymbol{\upsilon}_t$.  Analogously, we can correct the bias for $\boldsymbol{\omega}_t$ by using $\widetilde{\boldsymbol{\omega}}_t = \frac{\boldsymbol{\omega}_t}{(1-\beta_2)^t}$. 

Since $\widetilde{\boldsymbol{\omega}}_t$ denotes the moving average of bias corrected \textit{second} moment estimate, we use the ratio 
\begin{align}
    \frac{\widetilde{\boldsymbol{\upsilon}}_t}{\sqrt{\widetilde{\boldsymbol{\omega}}_t}} = \frac{\boldsymbol{\upsilon}_t}{\sqrt{\boldsymbol{\omega}_t}} \frac{\sqrt{1-\beta_2}}{1-\beta_1}.
    \label{eq:last}
\end{align}
Considering that $\boldsymbol{\upsilon}_t$ and $\boldsymbol{\omega}_t$ are vectors in Eq.~(\ref{eq:last}), we re-write the above as the following mathematically meaningful expression:
\begin{align}
   \boldsymbol{\rho} =  \frac{\widetilde{\boldsymbol{\upsilon}}_t}{\sqrt{\widetilde{\boldsymbol{\omega}}_t}} = \frac{\sqrt{1-\beta_2^t}}{1-\beta_1^t}~\text{diag}\left( \text{diag}(\sqrt{\boldsymbol{\omega}_t})^{-1} \boldsymbol{\upsilon}_t \right),
\end{align}
where diag(.) forms a diagonal matrix of the vector in its argument or forms a vector of the diagonal matrix provided to it. The inverse in the above equation is element-wise.   

\vspace{5mm}
\noindent{\bf A-2: Label leakage  suppression in LUTA}
\vspace{3mm}\\
To demonstrate the effect of leakage suppression using non-source classes, we perform the following experiment. In Algorithm~1, we replace the set $\overline{S}$ of the non-source classes with set $\mathcal{S}$ of the source class samples, and use `$\ell_{\text{target}}$' instead of `$\ell$' to compute the gradients. This removes any role the non-source classes play in the original LUTA.
Under the identical setup as for Table~\ref{tab:success} of the paper, we observe the following average percentage Leakage `rise' for the perturbations. VGG-16: 18.5\%, ResNet-50: 21.9\%, Inception-V3: 63.6\% and MobileNet-V2: 26.7\%. On the other hand, the changes in the fooling ratios on the test data are not significant. Concretely, the average test data fooling ratio changes are, VGG-16: $-1.2\%$, ResNet-50: $-0.4\%$, Inception-V3: $+5.2\%$, and MobileNet-V2: $-5.2\%$. Here, `$-$' indicates that the fooling ratio on the test set actually decreases when label leakage is not suppressed. Conversely, `$+$' indicates a gain, which occurs only in the case of Inception-V3. However, the label leakage rise for the same network is also the maximum. 
%of lower fooling ratio of the original LUTA for the Inception-V3. Although by giving multiple runs of the original stochastic algorithm or letting the algorithm run for longer can increase the test data fooling for Inception-V3, we strictly follow our protocols in Table~1 of the paper and report results of the first run. Hence, slightly better accuracy on Inception
This experiment conclusively demonstrates the successful label leakage suppression by the original algorithm. 

%{\color{red} TO WRITE}

\newpage
\noindent{\bf A-3: Label details for ImageNet model fooling}
\vspace{1mm}

For the labels used in Table~\ref{tab:success}  of the paper, Table~\ref{tab:Supp_T1} provides the detailed names and WordNet IDs.

\begin{table}[h]
    \centering
    {%\footnotesize
    \begin{tabular}{c|l|c}
    \hline
       Transformation & \multicolumn{1}{|c|}{ImageNet Label}  & WordNet ID \\ \hline \hline
       T$_1$ & source: airship, dirigible & n02692877\\
          & target: school bus & n04146614\\ \hline
       T$_2$ & source: ostrich, Struthio camelus & n01518878 \\
          & target: zebra & n02391049\\ \hline
       T$_3$ & source: lion, king of beasts, Panthera leo & n02129165 \\
          & target: orangutan, orang, orangutang, Pongo pygmaeus & n02480495 \\ \hline
       T$_4$ & source: bustard &    n02018795 \\ 
          & target: Arabian camel, dromedary, Camelus dromedarius & n02437312 \\ \hline
       T$_5$ & source: jellyfish &  n01910747 \\
          & target: killer whale, killer, orca, ..., Orcinus orca & n02071294 \\ \hline
       T$_6$ & source: lifeboat & n03662601 \\
          & target: great white shark, white shark, man-eater, ..., carc1harias & n01484850 \\ \hline
       T$_7$ & source: scoreboard & n04149813\\
          & target: freight car & n03393912 \\ \hline
       T$_8$ & source: pickelhaube & n03929855 \\
          & target: stupa, tope & n04346328 \\ \hline
       T$_9$ & source: space shuttle & n04266014 \\
          & target: steam locomotive & n04310018 \\ \hline
       T$_{10}$ & source: rapeseed  & n11879895 \\
          & target: sulphur butterfly, sulfur butterfly & n02281406 \\ \hline \hline
    \end{tabular}}
    \caption{Detailed labels and WordNet IDs of ImageNet for Table~\ref{tab:success} of the paper.}
    \label{tab:Supp_T1}
    \vspace{-5mm}
\end{table}

\vspace{3mm}
\noindent{\bf A-4: Further illustrations of perturbations for ImageNet model fooling}
\vspace{1mm}

\begin{figure}[h]
    \centering
    \includegraphics[width = \textwidth]{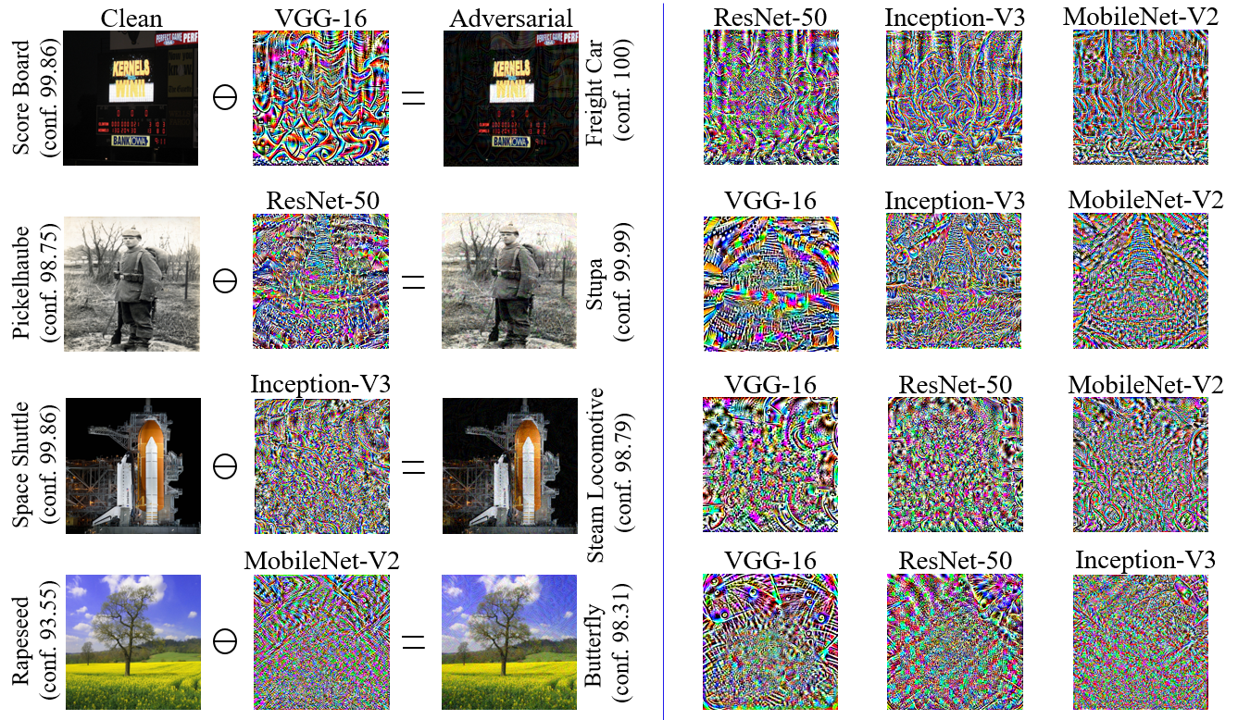}
    \caption{$\ell_{\infty}$-norm bounded perturbations with `$\eta = 15$'. A row contains perturbations for the same source $\rightarrow$ target fooling. Representative adversarial samples are also shown. We follow [1] for visualizing the perturbation. The perturbations are generally  hard to perceive for humans.}
    \label{fig:Linf_2}
\end{figure}

\begin{figure}[h]
    \centering
    \includegraphics[width = \textwidth]{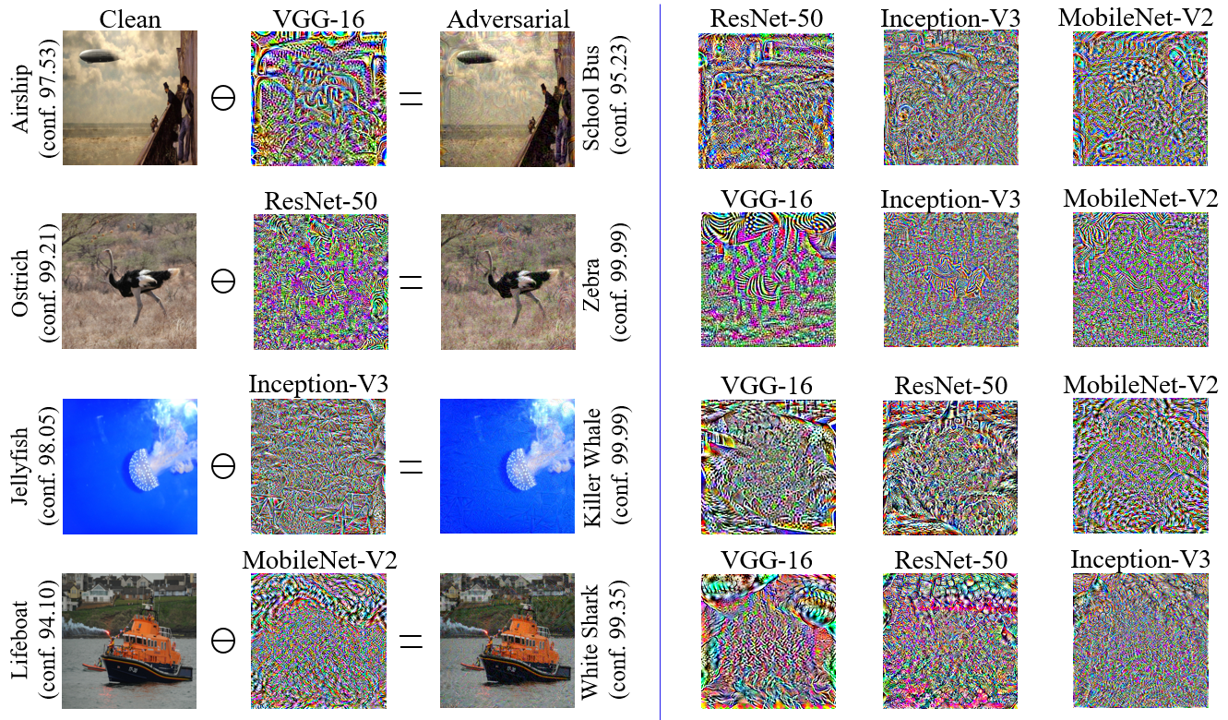}
    \caption{$\ell_{2}$-norm bounded perturbations with `$\eta = 4,500$'.}
    \label{fig:L2_1}
\end{figure}

\begin{figure}[h]
    \centering
    \includegraphics[width = \textwidth]{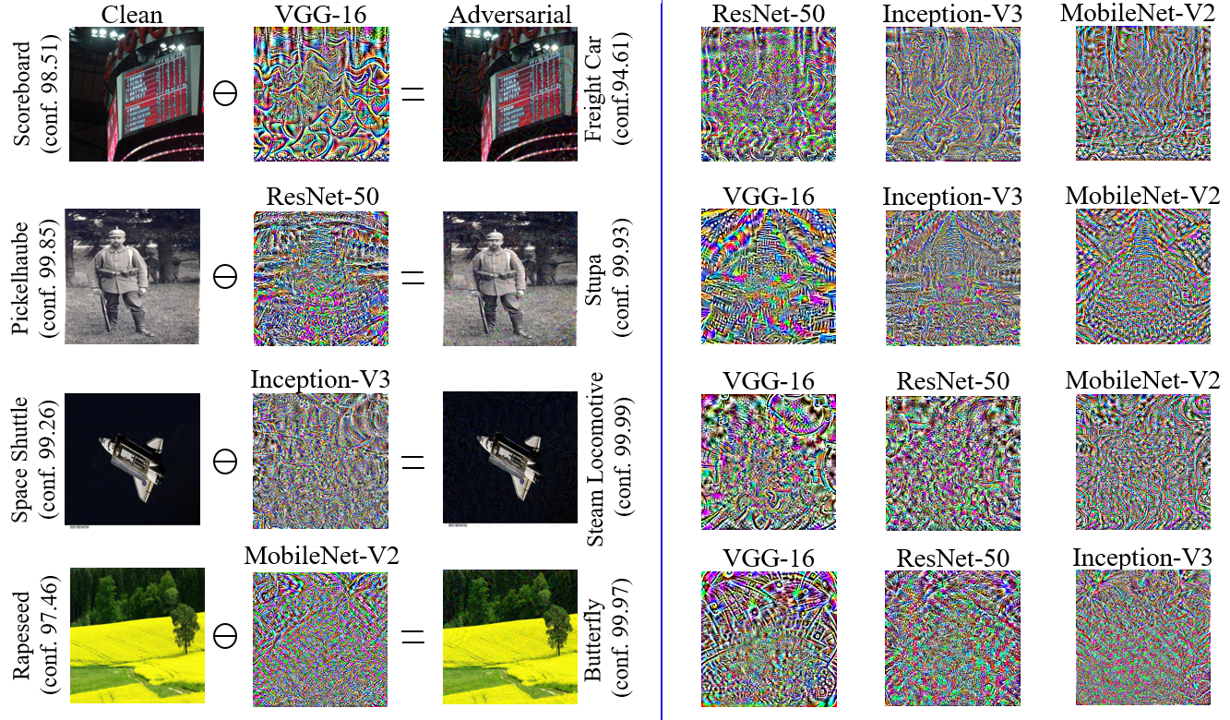}
    \caption{Further examples of $\ell_{2}$-norm bounded perturbations with `$\eta = 4,500$'.}
    \label{fig:L2_2}
\end{figure}

Fig.~\ref{fig:Linf_2} shows further examples of $\ell_{\infty}$-norm bounded perturbations with `$\eta = 15$'. We also show $\ell_2$-norm bounded perturbation examples in Fig.~\ref{fig:L2_1} and \ref{fig:L2_2}.

\vspace{5mm}

% \newpage
% \noindent{\bf A-5: Further results on transferability of LUTA:}\\
% In Table~\ref{tab:transfer2}.... 

% \begin{table}[]
%     \centering
%     \begin{tabular}{l||c|c|c} \hline%\toprule  
% Model & VGG & ResNet & Mob.Net \\\hline \hline
% VGG         &87.3$\pm$5.1 & 65.4$\pm$9.8 & 69.4$\pm$8.9\\  \hline %\midrule
% ResNet      &3 & 5 &\\  \hline %\midrule
% Mob.Net     &3 & 5 &\\ \hline %\bottomrule
% \end{tabular}
%     \label{tab:transfer2}
% \end{table}
\newpage
\noindent{\bf A-5: Further images of face identity switches:}
%The following images are computed for fooling VGGFace model. 

\begin{figure}[h]
    \centering
    \includegraphics[width = \textwidth]{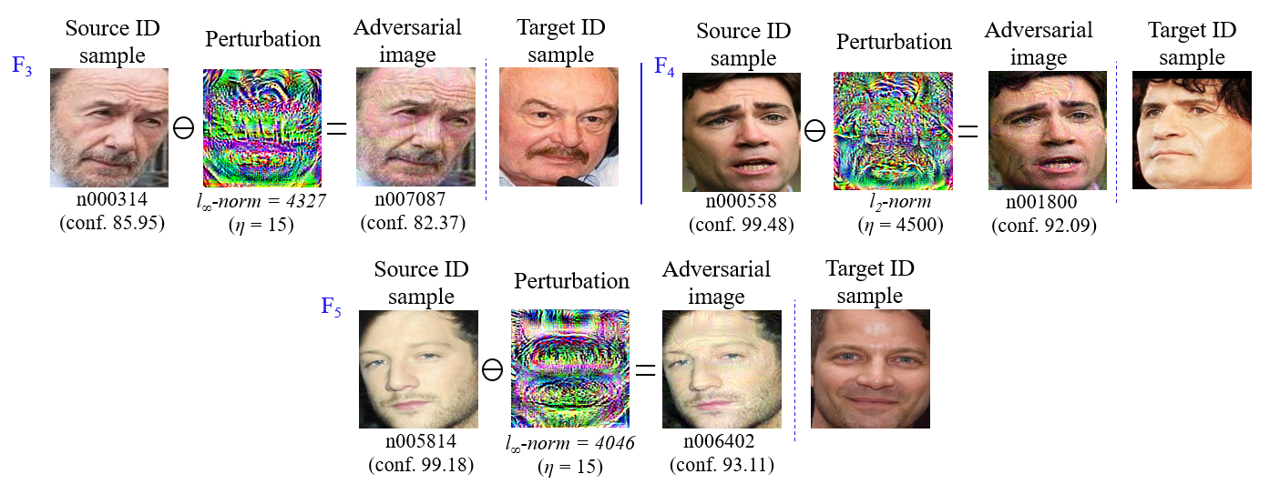}
    \vspace{-3mm}
    \caption{Representative  $\ell_{\infty}$ and $\ell_2$-norm bounded  perturbations for face identity switching on VGGFace model. Example clean images of the target classes are provided for reference only.}
    \label{fig:face2}
\end{figure}

\noindent{\bf A-6: Perturbation patterns:}

\begin{figure}[h]
    \centering
    \includegraphics[width = 0.8\textwidth]{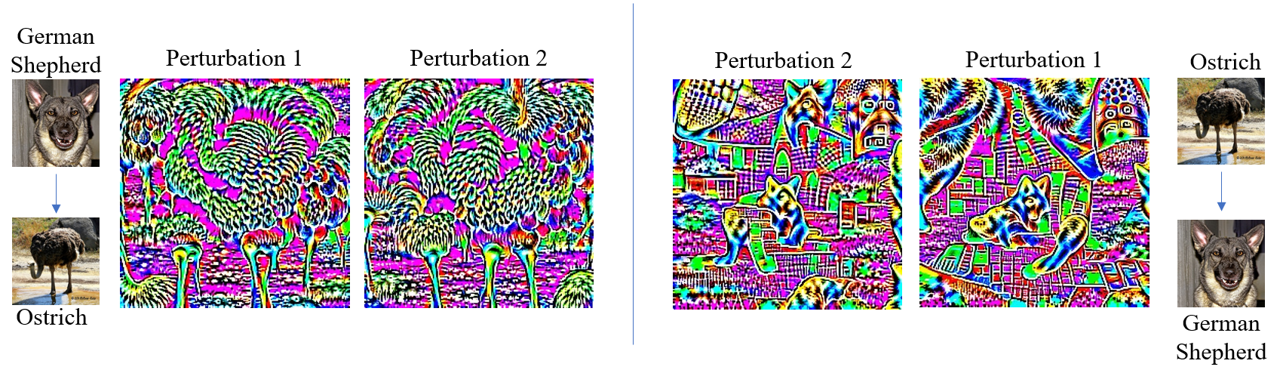}
    \caption{Multiple runs of LUTA result in different perturbation patterns. However, each patterns contains the dominant visual features of the target class. Clean samples are shown for reference only. }
    \label{fig:Multi_init}
    \vspace{-3mm}
\end{figure}

\begin{figure}[h]
    \centering
    \includegraphics[width = 3in]{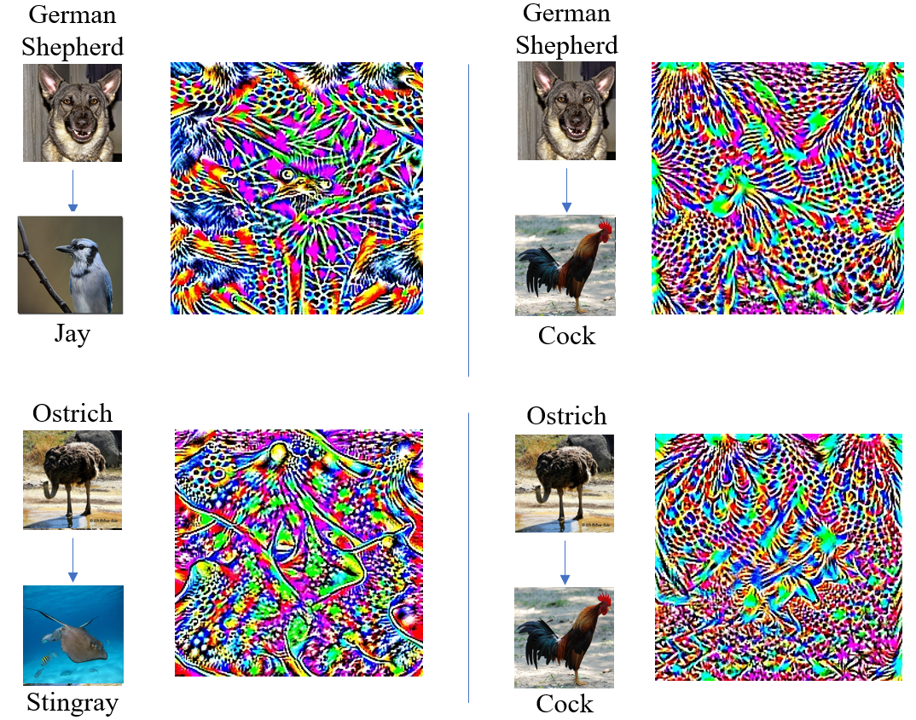}
    \caption{Further examples of perturbations for VGG-16. Distinct visual features of the target class are apparent in the perturbation patterns.}
    \label{fig:Others}
\end{figure}

With different LUTA runs for the same source $\rightarrow$ target transformations, we achieve different perturbations due to stochasticity introduced by the mini-batches. However, all those perturbations preserve the characteristic visual features of the target class. Fig.~\ref{fig:Multi_init} illustrates this fact. The shown perturbations are for VGG-16. We choose this network for more  clear target class patterns in the perturbations. This phenomenon is generic for the models. However, for more complex models, regularities are relatively harder to perceive. Fig.~\ref{fig:Others} provides few more VGG-16 perturbation examples in which visual appearance of the target classes are clear. 

\noindent{\bf A-7: LUTA in the Physical World:}
\vspace{2mm}

Label-universal targeted fooling has a challenging objective of mapping a large variety of inputs to a single (incorrect) target label. Considering that, a straightforward extension of this attack to the Physical World seems hard. However, our experiments demonstrate that label-universal targeted fooling is achievable in the Physical World using the adversarial inputs computed by the proposed LUTA.

To show the network fooling in the Physical World, we adopt the following settings. A $224 \times 224$ image (from ImageNet) is expanded to the maximum allowable area of A4-size paper in the landscape mode. We perform the expansion with a commonly used  image organizing software designed for personal photo management for the GNOME desktop environment, `Shotwell' (click \href{https://en.wikipedia.org/wiki/Shotwell_(software)}{here} for more details on the software). The software choice is random, and we prefer a common software because an actual attacker may also use something similar. After the expansion, we print the image on a plain A4 paper using the commercial  \href{https://www.konicaminolta.com.au/products/office-printing/office-multifunction-printers/bizhub-c458}{bizhub-c458} color printer from Konica-Minolta. Default printer setting is used in our experiments.
We use the same settings to print both clean and adversarial images.
The printed images are shown to a regular laptop webcam and its live video stream is fed to our target model that runs on Matlab 2018b using the `deep learning toolbox'. We use VGG-16 for this experiment.
We use a square $720\times 720$ grid for the video to match our square images. Note that, we are directly fooling a classifier here (no detector), hence the correct aspect ratio of the image is important in our case.

In the video we provide \href{https://www.youtube.com/watch?v=SiNLKbngBUI}{here}, it is clear that the perturbations are able to fool the model into the desired target labels quite successfully. For this experiment, we intentionally selected those adversarial images in which the perturbations were relatively more perceptible, as they must be visible to the webcam (albeit slightly) to take effect. Nevertheless, all the shown images use $\eta = 15$ for the underlying $\ell_{\infty}$-norm bounded perturbations. Perceptibility of the same perturbation can be different for different images, based on image properties (e.g.~brightness, contrast). For the images where the perturbation perceptibiltiy is low for the Physical World attack, a simple scaling of the perturbation works well (instead of allowing larger $\eta$ in the algorithm). However, we do not show any such case in the provided video. All the used image perturbations are directly computed for $\eta = 15$.  

It is also worth mentioning that a Physical World attack setup similar to [14] was also tested in our experiments, where instead of a live video stream, we classify digitally  scanned and cropped adversarial images (originally printed in the same manner as described above). However, for the tested images with quasi-imperceptible perturbations (with $\eta =15$), 100\% successful fooling was observed. Hence, that setup was not deemed  interesting enough to be reported. Our current setup is more challenging because it does not assume static, perfectly cropped, uniformly illuminated and absolutely plane adversarial images. These assumptions are implicit in the other setup.

\end{document}